\documentclass[aps,pra,reprint,twocolumn,superscriptaddress,amsmath,amssymb,citeautoscript]{revtex4-1}

\usepackage{graphicx}
\usepackage{bm,amsmath,amssymb}
\usepackage{mathrsfs}
\usepackage{dcolumn}
\usepackage{mathrsfs}
\usepackage{cases}
\usepackage{color}
 \usepackage{ulem}
 \usepackage{indentfirst}
\usepackage{epstopdf}
\usepackage{grffile}

\usepackage{adjustbox}
\usepackage{subfigure}

\DeclareGraphicsExtensions{.eps}

\newcommand{\be}{\begin{equation}}
\newcommand{\ee}{\end{equation}}
\newcommand{\bea}{\begin{eqnarray}}
\newcommand{\eea}{\end{eqnarray}}
\newcommand{\la}{\langle}
\newcommand{\ra}{\rangle}
\newcommand{\ii}{\rm i}

\begin{document}
\title{Supersolid and pair correlations of the extended Jaynes-Cummings-Hubbard model on triangular lattices}
\author{Lijuan Guo}
\affiliation{College of Physics and Optoelectronics, Taiyuan University of Technology, Shanxi 030024, China}
\author{Sebastian Greschner}
\affiliation{Department of Quantum Matter Physics, University of Geneva, 1211 Geneva, Switzerland}
\author{Siyu Zhu}
\affiliation{College of Physics and Optoelectronics, Taiyuan University of Technology, Shanxi 030024, China}
\author{Wanzhou Zhang}
\affiliation{College of Physics and Optoelectronics, Taiyuan University of Technology, Shanxi 030024, China}
\date{\today}
\begin{abstract}
We study the extended Jaynes-Cummings-Hubbard model on triangular cavity lattices and zigzag ladders.
By using density-matrix renormalization group methods, we observe various types of solids with different density patterns and find evidence for light supersolids, which exist in extended regions of the phase diagram of the zigzag ladder. Furthermore, we observe strong pair correlations in the supersolid phase due to the interplay between the atoms in the cavities and atom-photon interaction. By means of cluster mean-field simulations and a scaling of the cluster size extending our analysis to two-dimensional triangular lattices, we present evidence for the emergence of a light supersolid in this case also.
\end{abstract}
\maketitle

\section{Introduction}
\label{sec:intro}
Searching for novel supersolids~(SS)
and exploring their nature is an interesting topic in the field of condensed matter physics~\cite{BEC1,BEC2,BEC3,over}.
The controllable ultracold-atom system provides a pristine and convenient platform to realize such tasks~\cite{BEC4,BEC5} with multi-component systems of ultracold atoms being possible candidates to host the SS phase~\cite{p1,p2,p3,p4}. In particular Bose-Einstein-condensates in cavities have allowed for the successful observation of supersolids in experiments~\cite{Leonard2017a, Leonard2017b}.
Recently, supersolid properties have been found also in systems of dipolar quantum droplets~\cite{Tanzi2019,Boettcher2019}.

The Jaynes-Cummings-Hubbard~(JCH) model, a combination of the Jaynes-Cummings~(JC) model~\cite{QED,SMI} of coupled cavities where each cavity contains a two-level atom, is an interesting variant of cavity coupled quantum systems. Experimentally, the JCH model can be realized by a coupled transmission-line resonator~\cite{Low} or trapped ions~\cite{Experimental}.
Analytically, the mean-field~(MF) theory~\cite{MF, MF1}, the Ginzburg-Landau theory~\cite{GL}, the strong-coupling random-phase approximation method~\cite{strong} and the Green-function method~\cite{qpd}
are all used to study the properties of the JCH model.
Furthermore, the correlation and critical exponents of the JCH model
can be obtained by many reliable numerical methods, such as the density-matrix renormalization group algorithm (DMRG)~\cite{AM,Fermionized} and the quantum Monte-Carlo (QMC) method~\cite{Dynamical critical, zhaojize}.

Moreover, several interesting topics concerning the JCH model have been studied, which include fractional quantum Hall physics~\cite{Fractional}, quantum transport~\cite{Quantum transport}, quantum-state transmission~\cite{Quantum-state}, on-site disorder~\cite{MF1,Disorder}, three-body interactions~\cite{etb} and the interesting quantum phase transition between the superfluid~(SF) phase and the Mott-insulator~(MI) phase~\cite{SMI}. All of these previous works ignored the interaction between atoms. Until recently, the light supersolid was found in the Dicke model of a cavity modeled by quantum electrodynamics coupled with a one dimensional Rydberg lattice~\cite{zxf}.However, since the photon hopping between each cavity was not considered, it remains unclear whether or not the photon hopping will break the supersolid phase. At the same time, the regimes of the SS phase by a hole or particle-excited mechanism is very narrow in the phase diagram. In Ref.~\cite{Supersolid} a light supersolid in the extended JCH model on the square lattices was found by MF methods.

In the limit of a dominant atom-photon coupling, as will be
discussed below, the extended JCH model may be mapped on a Bose-Hubbard (BH) model. Since for the extended BH model, a broad regime of supersolid phases was found, in particular, on triangular lattices~\cite{tri2,tri3,pairss}, it is interesting to study the extended JCH model on the triangular cavity lattices away from this limit, and check whether additional SS exists in the phase diagram. Here the light supersolid phase may be stabilized by an order-by-disorder mechanism as discussed in Refs.~\cite{tri2,tri3,pairss}.

In this work, we study the extended JCH model on the triangular zigzag ladders and lattices in the low filling regime by means of arguments in limiting cases and detailed numerical DMRG and cluster mean field~(CMF) simulations. For the quasi-one-dimensional zig-zag ladder case we present example phase diagrams, which exhibit gaped density-wave (DW) phases at densities $\rho=1/3$, $1/2$ and $2/3$, as well as two extended regions of a pair-supersolid phase between the DW phases. We analyze properties of these phases and study the phase transitions to superfluid regions. We argue that the pairing inside the supersolid phase is a feature of the inherent quasi-1D geometry of the ladder system.
In order to extend our study to 2D-triangular lattice geometries, we employ cluster Gutzwiller mean-filed simulations. We find a qualitatively similar phase diagram with two gaped DW phases at $\rho=1/3$, and $2/3$
as well as extended supersolid phases in between. In relation to previous results in a related bosonic model ~\cite{tri2, tri3,pairss}, for which we present an effective mapping in a limiting case, the emergence of these supersolid phases may be explained by an order-to-disorder like mechanism. We present the qualitative general 2D phase diagram by MF simulations and show evidence at the stability of its main features, in particular of the SS and gaped phases by comparison to a simulation with various clusters sizes up to $12$ sites.
The experimental signatures of the SS phase are also shown by the momentum distribution and correlation.

The outline of this work is as follows. Section \ref{sec:model} shows the JCH model and discusses the MF and the DMRG methods used in the paper.
Section \ref{sec:zzJCH} shows the results of the JCH model on the triangular zigzag ladder by the DMRG method. In Section \ref{sec:two dimensional}, triangular lattices by the MF method are discussed.
Concluding comments are made in Section \ref{sec:con}.

\section{Model and Methods}
\label{sec:model}
\begin{figure}[tb]
\centering
\includegraphics[width=0.45\textwidth]{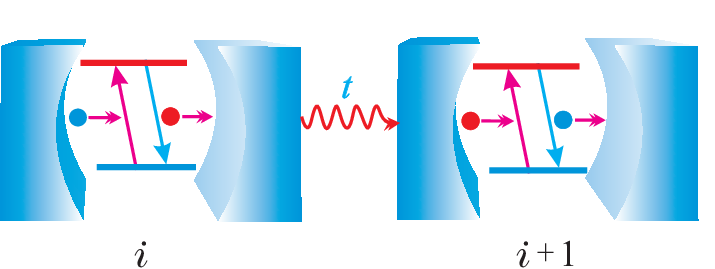}
\caption{A photon denoted by a red symbol is tunneling between two different cavities which are labeled by  $i$ and $i+1$, and $t$ is the hopping strength. In each cavity, the atom has two energy levels which are labeled by two separated horizontal lines.}
\label{fig:ex}
\end{figure}

Following Ref.~\cite{Supersolid}, we consider a triangular lattice of coupled cavities as sketched in Fig.~\ref{fig:ex}. On each cavity site $i$, the two-level atom with a ground state $|g\ra$ and excited state $|e\ra$ is contained.
The on-site coupling between the photons and the atom on each site $i$ can be described by the JC Hamiltonian $H_i^{JC}$
\begin{equation}
\begin{aligned}
H_i^{JC}=&\omega n_i^a+\varepsilon n_i^\sigma+\beta (a_i^\dagger \sigma_i+a_i \sigma_i^\dagger),
\label{b}
\end{aligned}
\end{equation}
where $\omega$ is the frequency of the mode of the photon creation and annihilation operators at lattice site $i$,
$\varepsilon$ is the transition frequency between two energy levels, $n_i^a=a_i^\dagger a_i$ and $n_i^\sigma=\sigma_i^\dagger \sigma_i$
 are the photon number and the number of excitations of the atomic levels, respectively. $a_i^\dagger$ and $a_i$ are respectively the photon creation and annihilation operators at lattice site $i$. The pauli matrices
$\sigma_i^\dagger$ ($\sigma_i$) represent the raising (lowering) operator, and $\beta$ is the atom-photon coupling  strength.
As shown in Fig.~\ref{fig:ex}, $a_i \sigma_i^\dagger$ means that a photon is absorbed and an atom excitation forms simultaneously, and may hence be understood as a raising operator in a pseudo-spin -1/2 space formed by the two states $|g,1\rangle_x$ and $|e,0\rangle_x$.
For convenience, we focus on the case $\varepsilon=\omega=0$.

In this study, we focus our analysis to a sector of low density of excitations $\rho \leq 1$, where at most one photon is present in each cavity. As we will show, this regime can be modeled by a hard-core particle description with at most one photon per cavity.

\begin{figure}[b]
\centering
\includegraphics[width=0.45\textwidth]{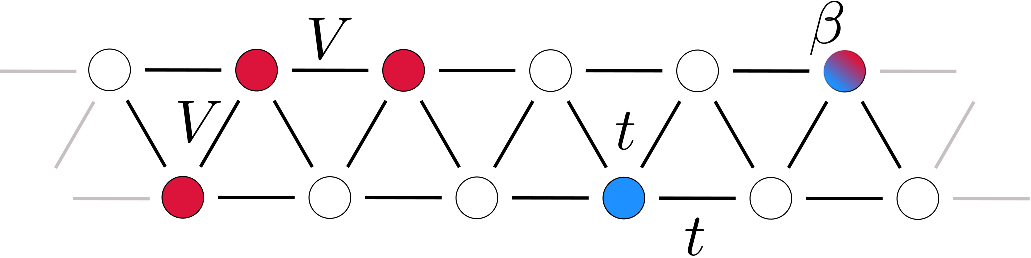}
\caption{Sketch of the extended JCH model on a  triangular cavity ladder geometry.}
\label{fig:sketch}
\end{figure}

The extended JCH model includes a dipole interaction term and photon tunneling term between cavities. The Hamiltonian is defined as
\begin{equation}
\begin{aligned}
H=&\sum_{i}(H_i^{JC}-\mu n_i)-t\sum_{\la i,j\ra}(a_i^\dagger a_j+\rm H.c.)\\
&+\sum_{\la i,j\ra} V n_i^\sigma n_j^\sigma,
\label{a}
\end{aligned}
\end{equation}
where the total number of excitations is $\rho \equiv \sum_i{n}_i=\sum_i(n_i^\sigma+n_i^a)$, $\mu$ is the chemical potential, $t$ is the  hopping amplitude of  photons between a pair of neighboring lattice sites $i$ and $j$,  and $V$ is the nearest-neighbor interactions between the atoms.
The main ingredients of model~\eqref{a} are sketched in Fig.~\ref{fig:sketch} for a zigzag-ladder geometry.


\subsection{Limit of dominant atom-photon coupling}

In the limit of a dominant atom-photon coupling $\beta\gg V$ and $t$ when fillings $\rho<1$, one may project model~\eqref{a} to its low energy subspace composed of sites with on-site singlets $(|g,1\rangle_x-|e,0\rangle_x)/\sqrt{2}$ and empty sites $|g,0\rangle_x$, after identification of the states $|1\rangle^b_x$ and $|0\rangle^b_x$ in a (hardcore) BH model. Hence, a first order approximation map model~\eqref{a} is given by the following Hamiltonian
\begin{equation}
\begin{aligned}
H_{BH}=&\sum_{i}(-\beta-\mu) n_i^b -\frac{t}{2}\sum_{\la i,j\ra}(b_i^\dagger b_j+{\rm H.c.})\\
&+\frac{V}{4}\sum_{\la i,j\ra} n_i^b n_j^b,
\label{aBH}
\end{aligned}
\end{equation}
with bosonic annihilation (creation) operators $b_i$ ($b_i^\dagger$) and $n_i^b=b_i^\dagger b_i$.
The properties of model~\eqref{aBH} have been studied extensively in various lattice geometries, for example, Refs.~\cite{tri2, tri3,pairss, Mishra14, Mishra15}. In the following, we focus on the properties of the model~\eqref{a} in the regime $t<V\lesssim \beta$.


\section{Extended JCH model on triangular zigzag ladders}
\label{sec:zzJCH}

In the following, we study the JCH model on quasi-1D zigzag ladders.

\subsection{Limit of vanishing inter-cavity hopping}

\begin{figure}[b]
\centering
\begin{minipage}[t]{.1\linewidth}\raisebox{1.1\height}{(a)}\end{minipage}%
\begin{minipage}{.9\linewidth}
\includegraphics[width=\linewidth,trim={6cm 0 6cm 0},clip]{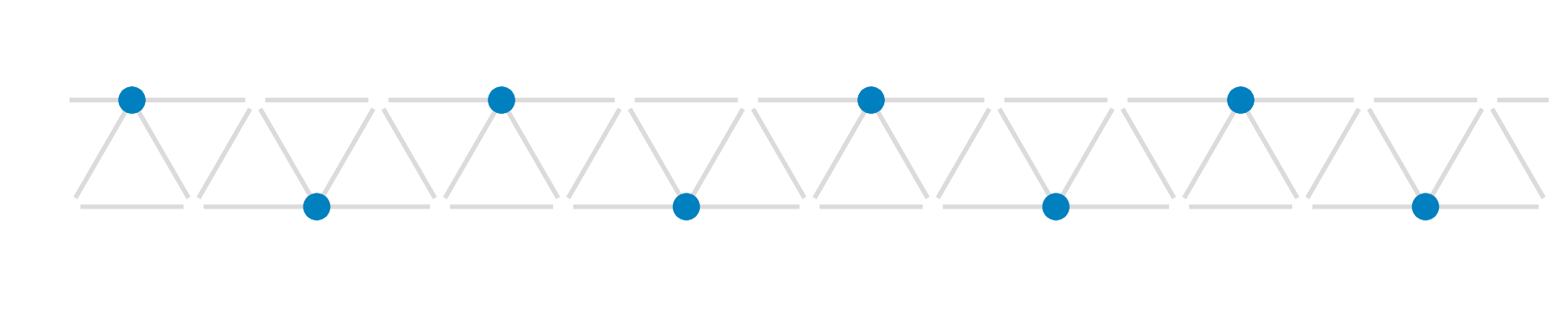}
\end{minipage}\\[-0.5cm]
\begin{minipage}[t]{.1\linewidth}\raisebox{1.1\height}{(b)}\end{minipage}%
\begin{minipage}{.9\linewidth}
\includegraphics[width=\linewidth,trim={6cm 0 6cm 0},clip]{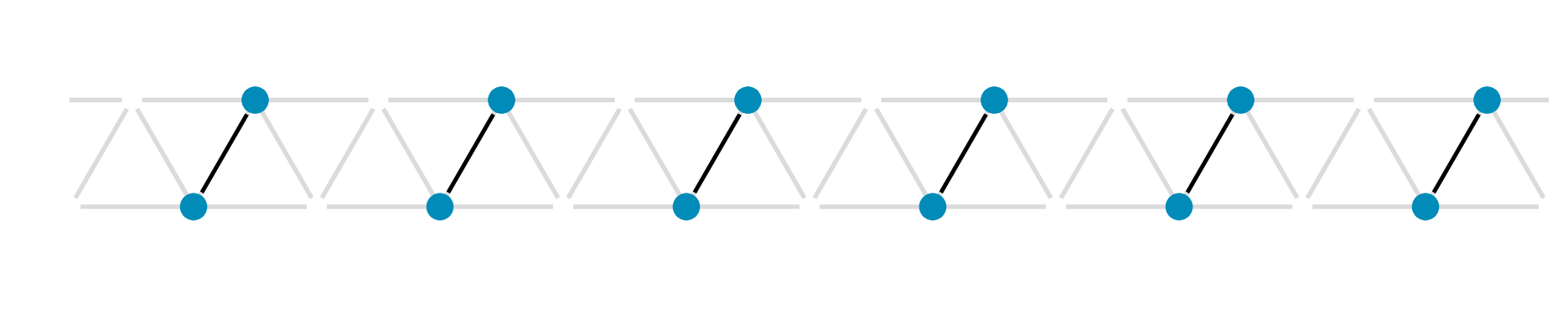}
\end{minipage}\\[-0.5cm]
\begin{minipage}[t]{.1\linewidth}\raisebox{1.1\height}{(c)}\end{minipage}%
\begin{minipage}{.9\linewidth}
\includegraphics[width=\linewidth,trim={6cm 0 6cm 0},clip]{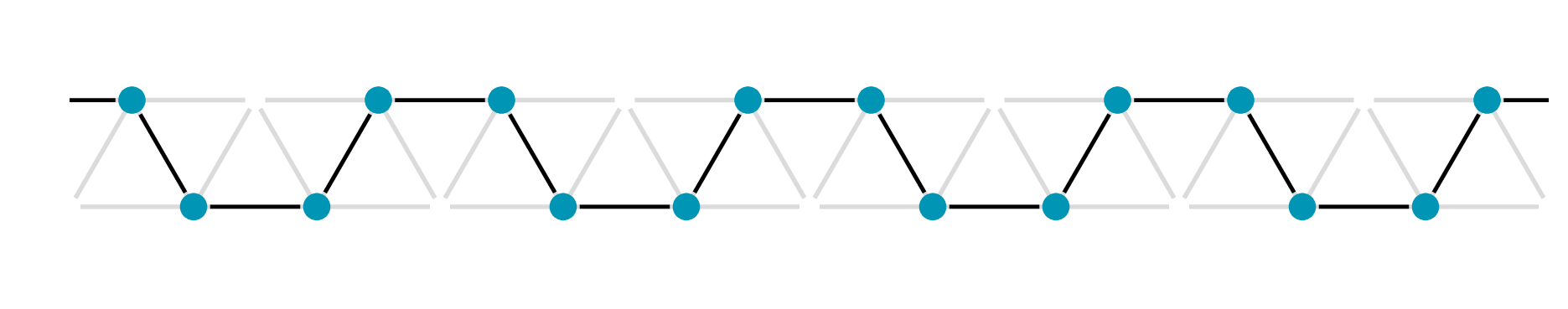}
\end{minipage}\\[-0.5cm]
\caption{Sketch of the typical density configurations (blue bullets) for the ladder for the (a) DW$_{1/3}$ phase ($t/\beta=0.015$, $\mu/\beta=-0.9$) and the (b) DW$_{2/3}$ phase ($t/\beta=0.015$, $\mu/\beta=-0.7$). The black lines indicate the strength of the nearest-neighbor density-correlations $\la n^b_i n^b_j \ra$.}
\label{fig:zz_phases_sketch}
\end{figure}
\begin{figure}[t]
\includegraphics[width=\linewidth]{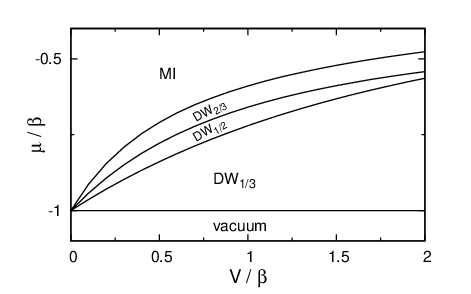}
\caption{Phase diagram of the extended JCH model on a quasi-1D zigzag ladder (DMRG-simulations) in the limit of vanishing cavity coupling $t=0$, as discussed in the text, as a function of the chemical potential $\mu$ and the nearest-neighbor interaction $V$.}
\label{fig:fig_zzJCH_pdt0_beta1_V0.4}
\end{figure}

The limit of a vanishing tunneling $t\to 0$ the ground states of the (extended) BH model such as Eq.~\eqref{aBH} are given by classical particle configurations. At a fixed filling, the ground-state manifold is yet typically highly degenerate as excitations are localized at some lattice site. For certain filling fractions, e.g. on zigzag ladder $\rho=1/3$, $\rho=1/2$ and $\rho=2/3$, the degeneracy is lifted, where a DW pattern minimizes the interaction energy $V$. Therefore, only those filling fractions remain stable in a grand-canonical ensemble. The lattice configurations of the extended JCH model in the $t\to 0$ limit resemble the ones of the extended BH model physics with stable plateaus at fillings $\rho=1/3$, $\rho=1/2$ and $\rho=2/3$, and a macroscopic degeneracy for the remaining fillings. We sketch the configurations in Fig.~\ref{fig:zz_phases_sketch}.
An important difference between the quasi-1D ladder geometry and the 2D lattices is that in the triangular two-leg ladder, a further solid density wave phase at half-filling can be found corresponding to a pattern $(0,0,1,1)$ in addition to the solid phases at fillings $1/3$ and $2/3$. The emergence and properties of these solid phases for the BH limit have been studied in various works, e.g. Refs.~\cite{Mishra14,Mishra15}.

Due to the interplay between $\beta$ and $V$, the JCH model, however,
obviously exhibits a nontrivial physics in the pseudo-spin sector even for  strict $t = 0$ limit. The $V$ term leads to Ising-like interactions of excitation on neighboring sites. Expressing the states $|g,1\rangle_j$ as a spin $|\downarrow\ra_j$ and $|e,0\rangle_j$ as $|\uparrow\ra_j$, we may express the Hamiltonian of the pseudo-spin degree of freedom as an Ising model with a transverse and longitudinal field
\begin{align}
H_{1/2} = V \sum_{\la j,j'\ra} \left(S^z_j-\frac{1}{2}\right) \left(S^z_{j'}-\frac{1}{2}\right)  +  \beta \sum_j S^x_j,
\label{eq:Hising}
\end{align}
where the sum $\la j,j'\ra$ runs over neighboring occupied sites.
The DW$_{1/3}$ and  DW$_{1/2}$ phases form pseudo-spin singlets of single occupied or neighboring site occupations, while interestingly the DW$_{2/3}$ phase exactly maps to the 1D Ising model. As model~\eqref{eq:Hising} is non-integrable for $\beta\neq 0$, we calculate the ground-state energies by means of DMRG~\cite{DMRG1,DMRG2} simulations and obtain the generic phase diagram Fig.~\ref{fig:fig_zzJCH_pdt0_beta1_V0.4} of the JC model on the zigzag ladder.

\subsection{DMRG results}

In the following, we study the JCH model on quasi-1D zigzag ladders at finite $t\geq 0$ and intermediate strength interactions $V\lesssim \beta$. We employ DMRG simulations with open boundary conditions keeping up to $m=1000$ matrix states in the sector of a fixed number of excitations $\rho$.  We calculate several observables and correlation functions to characterize the various ground-state phases.

\subsubsection{The phase diagram}

\begin{figure}[t]
\includegraphics[width=\linewidth]{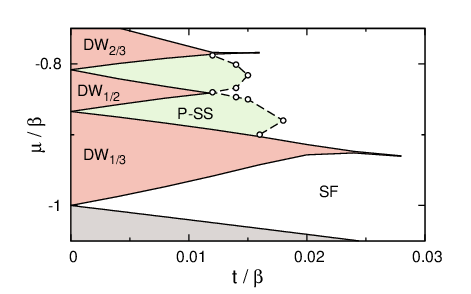}
\caption{Phase diagram of the extended JCH model on a quasi-1D zigzag ladder (DMRG-simulations, $V/\beta=0.4$). We observe several gaped DW phases at fillings $1/3$, $1/2$ and $2/3$ as well as a MI region at unit filling (not shown). The green shaded regions mark
the pair-supersolid (P-SS) surrounding the DW$_{1/2}$ phase (see text).}
\label{fig:fig_zzJCH_pd_beta1_V0.4}
\end{figure}
\begin{figure}[t]
\includegraphics[width=\linewidth]{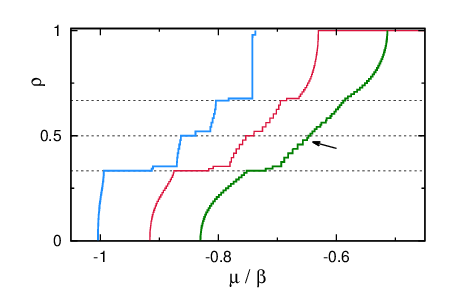}
\caption{$\mu-\rho$ curve for the extended JCH model on a quasi-1D zigzag ladder (DMRG-simulations, $V/\beta=0.4 $, $L=96$ sites) for different values of $t/\beta = 0.002$, $t/\beta = 0.008$ and $t/\beta = 0.015$ (from left to right). The curves have been shifted by $0.1$ among each other for clarity. The paired phase can be seen in this diagram by the presence of steps of $\Delta N=2$. The arrow marks the transition to the SF phase with $\Delta N=1$ for the $t/\beta = 0.015$ curve.}
\label{fig:fig_zzJCH_mag_beta1_V0.4}
\end{figure}

In Fig.~\ref{fig:fig_zzJCH_pd_beta1_V0.4}, we present the phase diagram of the extended JCH model on the zigzag ladder anticipating the discussion of the following section in the $\mu-t$ plane for $V=0.4\beta$ and for small fillings $\rho<1$. A finite hopping $t>0$ will generally destabilize the gaped phases and induces a transition of a gapless phase, which is an ordinary single component (superfluid) Luttinger-liquid like phase of photons tunneling between the cavities (SF). Interestingly, we also find an extended region around the DW$_{1/2}$ lobe, which we call a pair-supersolid phase (P-SS), which will be discussed below in more detail.

Fig.~\ref{fig:fig_zzJCH_mag_beta1_V0.4} shows the equation of state $\rho=\rho(\mu)$ for several cuts through the phase diagram of Fig.~\ref{fig:fig_zzJCH_pd_beta1_V0.4}. The dashed horizontal lines depict the commensurate fillings $\rho=1/3$, $1/2$ and $2/3$ for which we observe the gaped density wave phases, characterized by a plateau in the $\mu-\rho$ curve and a vanishing compressibility. Note, that due to finite size effects and the conservation of the total particle number $N=N_a+N^\sigma$, the $\mu-\rho$ curve is not smooth but consists of small steps $\Delta N$ and we observe a splitting of the plateaus into two levels at filling of $N$ particles on $L$ sites as well as $N+1$ or $N+2$.

\begin{figure}[t]
\includegraphics[width=\linewidth]{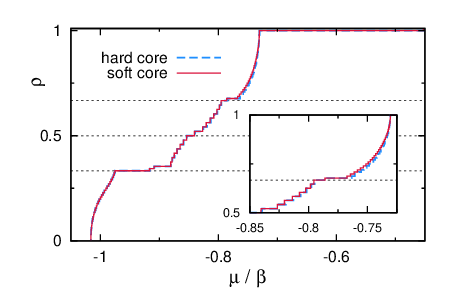}
\caption{$\mu-\rho$ curve for the extended JCH model on a quasi-1D igzag ladder (DMRG-simulations, $V/\beta=0.4$, $L=96$ sites, $t/\beta = 0.008$) for the hard-core and the soft-core JCH model (allowing for the occupation of 2 photons per cavity). Within the given parameters both curves coincide almost perfectly.}
\label{fig:fig_zzJCH_mag_beta1_V0.4_soft}
\end{figure}

While for the previous results we studied the hard-core JCH model, experimentally of course, the presence of more than one photon per cavity cannot be excluded. However, as can be seen in Fig.~\ref{fig:fig_zzJCH_mag_beta1_V0.4_soft}, this effect plays a minor role in the low density regime studied in this work. Relaxing the hard-core constraint only leads to a minor shift in the $\rho-\mu$ curves of Fig.~\ref{fig:fig_zzJCH_mag_beta1_V0.4_soft}.

\subsubsection{Pairing}

The paired phase can be identified in the $\mu-\rho$ curve of Fig.~\ref{fig:fig_zzJCH_mag_beta1_V0.4} by the presence of steps of $\Delta N=2$, compared to the ordinary SF-phase with steps $\Delta N=1$. This feature is typical for phases of paired particles, where the two-particle excitation energy $\Delta E_2$ becomes lower than the $\Delta E_1$ which dues to a gain in binding energy. Here we may define
\begin{align}
\Delta E_{\nu} = \frac{E(N-\nu) - 2 E(N) + E(N+\nu)}{\nu},
\end{align}
with the ground-state energy $E(N)$ of a system of $N$ excitations on $L$ sites.

\begin{figure}[b]
\centering
\begin{minipage}{.48\linewidth}
\includegraphics[width=\linewidth]{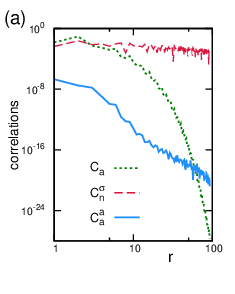}
\end{minipage}
\begin{minipage}{.48\linewidth}
\includegraphics[width=\linewidth]{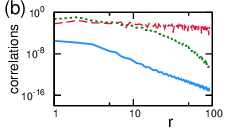}
\includegraphics[width=\linewidth]{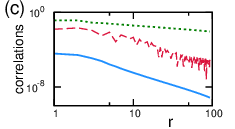}
\end{minipage}
\caption{
DMRG method detailed description of the extended JCH model on the triangular zigzag ladder with $V/\beta=0.4$, $\rho=0.416$ ($L=192$ sites) for (a)  $t/\beta=0.004$ (P-SS phase), (b)  $t/\beta=0.012$ (P-SS) and (c)  $t/\beta=0.026$ (SF). While $C_a^{\sigma}(r)$, $C_n^{\sigma}(r)$ show always a clear algebraic decay, the single particle correlation $C_a(r)$ is only algebraic in the SF phase, while exponentially in the P-SS phases.
}
\label{fig:fig_zzJCH_corr_beta1_V0.4_n0.416}
\end{figure}
In order to further characterize the P-SS and SF phases, we calculate correlation functions.
In previous both Luttinger-liquid phases,
the atom excitation correlation $C_n^\sigma(r)$ and the photon density correlation $C_n^a(r)$ decay with an inverse power-law~\cite{lt1,lt2,8},
where $C_n^\sigma(r)$ and $C_n^a(r)$ are defined by
{
\begin{equation}
\begin{split}
C_n^\sigma(r)&=\langle n_i^{\sigma}n_{i+r}^{\sigma}\rangle-\langle n_i^{\sigma}\rangle\langle n_{i+r}^{\sigma}\rangle,\\
C_n^a(r)&=\langle n_i^{a}n_{i+r}^{a}\rangle-\langle n_i^{a}\rangle\langle n_{i+r}^{a}\rangle.
\end{split}
\end{equation}}
At incommensurate fillings in general, the above correlations emerge in the shape of beats~\cite{6,beats}.
The superfluid order could be denoted by non-integer fillings and power-law decaying of the non-diagonal correlation\cite{lt1,lt2,8}
{
\begin{equation}
\begin{split}
C_a(r)&=\langle a_i^{\dagger}a_{i+r}\rangle.\\
\end{split}
\end{equation}}

Besides the usual single particle tunneling correlation, pairing-correlations
may be defined as
\begin{equation}
\begin{split}
C_a^\sigma(r)&=\langle a_i^{\dagger} \sigma_i^{\dagger} \sigma_{i+r} a_{i+r}\rangle,\\
C_a^a(r)&=\langle a_i^{\dagger} a_{i+1}^{\dagger} a_{i+r} a_{i+r+1}\rangle,\\
C_\sigma^\sigma(r)&=\langle \sigma_i^{\dagger} \sigma_{i+1}^{\dagger} \sigma_{i+r} \sigma_{i+r+1}\rangle.
\end{split}
\end{equation}

In Fig.~\ref{fig:fig_zzJCH_corr_beta1_V0.4_n0.416} we plot the correlations for a fixed density for various tuning rates. While in the SF phases, the density-density correlation $C_n$, the superfluid single particle $C_a$ and the pairing correlations $C_a^a(r)$ exhibit an algebraic decay, upon entering the P-SS region, the single particle correlations $C_a$  decay faster than a
 power law. The pairing correlations decay fast, but clearly with
an  power law (corresponding to a line in the log-log plot). For small tunneling rates (Fig.~\ref{fig:fig_zzJCH_corr_beta1_V0.4_n0.416}~(a)) the pairing correlations are also in absolute value exceeding the single-particle correlations.

Indeed, the presence of the pairing phase for the zigzag ladder may be understood already from the BH limit for a strong atom-photon coupling due to the presence of the solid phase at half filling $\rho=1/2$ for $t\rightarrow 0$. In this limit, we  observe that it is energetically favorable in a grand-canonical ensemble to dope
the system with an even number of holes such that the total size of domain-wall excitations (in pairs of 3 lattice sites) becomes
commensurate with the original crystalline lattice structure (4 lattice sites unit-cell). Hence, we may understand the dominant pairing correlation observed numerically in the JCH-model on a zigzag ladder as a reminiscent of this phase.
For a minimal handwaving example, one may analogously consider the doping of a single excitation on top of the DW$_{1/3}$, in a schematic picture given by $|\cdots -00-00-00-00-00\cdots \ra$, using the nomenclature of Eq.~\eqref{aBH}. In the limit of $\beta\gg V$ and $\beta\gg V$ also, an additional particle creates two domain-wall like excitations $|\cdots --0-00-00-00-00\cdots \ra$ at a cost of $\sim 2V/4 - \beta - \mu$, which can move through the lattice gaining kinetic energy, e.g. $|\cdots --00-000-00--00\cdots \ra$. Now a second photon can be added, e.g $|\cdots --00-00--00--00\cdots \ra$, at a lower cost $\sim 3 V/8 - \beta - \mu$.
The soft-core character of the JCH model allows to further lower the energy cost and thus maintain the preservation of the crystalline lattice structure, leading to the formation of a supersolid ordering, as will be shown in the following.

\subsubsection{Emergence of the supersolid correlations}

\begin{figure}[tb]
\includegraphics[width=0.48\linewidth]{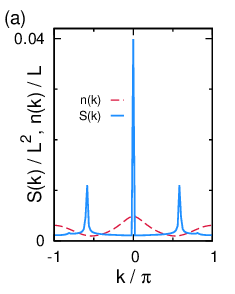}
\includegraphics[width=0.48\linewidth]{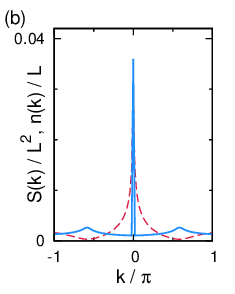}
\caption{
The structural factor $S(k)$ and momentum distribution $n(k)$ for the zigzag ladder for $\rho=5/12$, $V/\beta=0.4$ and (a) $t/\beta=0.008$, P-SS phase (b) $t/\beta=0.026$ SF phase ($L=96$ sites, DMRG simulation).  For (a), in the P-SS
 the third local maximum of the structural factor is also observed.}
\label{fig:zzstructurefactor}
\end{figure}

Interestingly, the P-SS region not only exhibits a strong pairing, but we also find evidence for a supersolid-like ordering. For this, we study the scaling of the structural factor which helps to characterize the solid order~\cite{3,4,5,6,7, Mishra2015,Greschner2017}
\begin{equation}
S(k)^{\nu} = \sum_{j, j'} \la n_j^\nu n_{j'}^\nu \ra \exp(\ii k (j-j') ),
\label{eq:Sk}
\end{equation}
where $\nu=\sigma, a$ and both quantities can reflect the solid order.
Apart from the trivial peak at $k=0$, $S(k)$ has a second pronounced
peak at $k_{max}=\pm 2/3\pi$ for the gaped DW phase. Also in the
gapless P-SS phase,
we observe a pronounced second maximum $S(k_{max})$ which is shifted
slightly away from $k_{max}=\pm 2/3\pi$.
Examples for the structure factor are shown in Fig.~\ref{fig:zzstructurefactor}(a). This peak indicates the presence of a density ordering in the liquid phase, which defines a supersolid.
In Fig.~\ref{fig:zzstructurefactor}, we also plot the Fourier transform of the $C_a(r)$, namely, the momentum distribution is defined as
\begin{equation}
n(k)=\frac{1}{L}\sum_{j,j'}\langle a_j^{\dagger}a_{j'}\rangle \exp (\ii k (j-j')),
\label{k}
\end{equation}
which was observed experimentally~\cite{BEC4}.
The exponential suppression of the single-particle correlations may also be seen by the blurring of the momentum-distribution in the P-SS phase.

\begin{figure}[tb]
\includegraphics[width=0.45\linewidth]{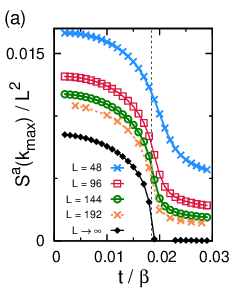}
\includegraphics[width=0.45\linewidth]{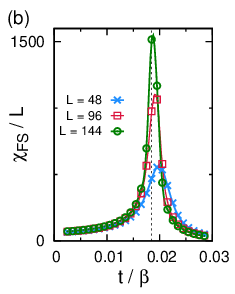}
\caption{Cuts through the phase diagram of Fig.~\ref{fig:fig_zzJCH_pd_beta1_V0.4} for a fixed density $\rho=5/12$. (a) DW-order in the SS-phase. Scaling of the peak maximum of the structure-factor in Eq.~\eqref{eq:Sk}  $S(k_{max})$ for different system-sizes $L=48, 96, 144$ and $192$. The lower black line is the extrapolation to the thermodynamic limit using a higher order polynomial. (b) Scaling of the fidelity susceptibility $\chi_{FS}/L$ according to Eq.~(\ref{eq:fidelity}) across the P-SS to SF phase transition.
}
\label{fig:fig_zzJCH_cuts_beta1_V0.4_n0.416}
\end{figure}

Following Refs.~\cite{3,4,5,7}, we study the scaling of the peak height of the structural factor $S(k_{max})$ with the system size. As shown in Fig.~\ref{fig:fig_zzJCH_cuts_beta1_V0.4_n0.416}~(a),  an extrapolation to the thermodynamic limit of the maximum structural factor $S(k_{max})$ is performed with different sizes for a cut through the phase diagram of Fig.~\ref{fig:fig_zzJCH_pd_beta1_V0.4}. The structure factor $S(k_{max})$ remains finite in the P-SS phase and vanishes after the transition to the SF region.

The phase transition point may be further classified by a study of the fidelity susceptibility~\cite{Fidelity}
\begin{equation}
\label{eq:fidelity}
\chi_{FS}(t) = \lim_{\delta t\to 0} \frac{-2 \ln |\langle \Psi_0(t) | \Psi_0(t + \delta t) \rangle| }{(\delta t)^2} \,,
\end{equation}
with the ground-state wave function $|\Psi_0 (t)\rangle$, a tunneling parameter $t$ and a small parameter $\delta t$.
In Fig.~\ref{fig:fig_zzJCH_cuts_beta1_V0.4_n0.416}~(a) we observe the emergence of distinct divergent peaks of $\chi_{FS}/L$ close to the phase transition between the P-SS and SF phase. The roughly linear scaling of $\chi_{FS}/L$ with the system size hints at an Ising character of the phase transition between the two regimes.

\section{CMF results for triangular lattices}
\label{sec:two dimensional}

\subsection{Cluster mean-field method}

The single-site MF has successfully predicted the SF-MI phase transition without long-range interaction ($V$=$0$)~\cite{opt}.
The cluster mean-field~(CMF) will be more reliable  in predicting the physics in the interaction systems ($V$$\ne$$0$)~\cite{cluster2, cluster3, cluster4, cluster5}.
The basic idea is to divide the system into $N_c$ unit cells, and each unit cell contains $nc$ sites.
The Hamiltonians within each cell are treated exactly and the Hamiltonians between each cell are
approximated by
$AB\approx A\la B \ra +\la A \ra B-\la A\ra \la B \ra$.

The total Hamiltonian can be considered as a sum over the local Hamiltonians on each unit cell, which contain the parts $H_{in}^c$, which are treated
 exactly, and the CMF Hamiltonian $H_{MF}^c$ as follows
\begin{equation}
\begin{aligned}
H=\sum_{c=1}^{N_c}(H_{in}^{c} +H_{MF}^{c}).
\label{d}
\end{aligned}
\end{equation}
The Hamiltonian $H_{in}^{c}$ can be expressed as
\begin{equation}
\begin{aligned}
H_{in}^{c}=&-zt\sum_{i,j\in c}(a_i^\dagger a_j+H.c.)+z V\sum_{i,j\in c}n_i^\sigma n_j^\sigma+\sum_{i\in c}h_i,
\label{f}
\end{aligned}
\end{equation}
where $h_i=-\mu_p n_i^a-(\Delta+\mu_s)n_i^\sigma+\beta(\sigma_i^\dagger a_i+a_i^\dagger \sigma_i)$.
The chemical potential of photons is $\mu_p$ and the chemical potential of atoms is $\mu_s$, and  the different labels of chemical potential are convenient to test our
codes. In real simulations, $\mu_p=\mu_s=\mu-\omega$, $\Delta=\omega-\varepsilon$, and $\Delta$ are maintained at zero for convenience.

The Hamiltonian $H_{MF}^{c}$ is given by
\begin{equation}
\begin{aligned}
H_{MF}^{c}=&-q t \sum_{i,j\in ce}[(a_i^\dagger+a_i) \Psi_j+(a_j^\dagger+a_j)\Psi_i-2\Psi_i \Psi_j]\\
&+q V\sum_{i,j\in ce}({n_i^\sigma \rho_j^\sigma+n_j^\sigma \rho_i^\sigma-\rho_i^\sigma \rho_j^\sigma }),
\label{g}
\end{aligned}
\end{equation}
where $z$ is equal to $1$, the chosen setting in Ref.~\cite{Supersolid},
$q=2z$ for the triangular lattices,
$\Psi_i=\la a_i\ra$ is the superfluid order parameter, and
$\rho_i^\sigma=\la n_i^\sigma\ra$ is the number of atomic excitations.

\begin{figure}[tb]
\centering
\includegraphics[width=0.45\textwidth]{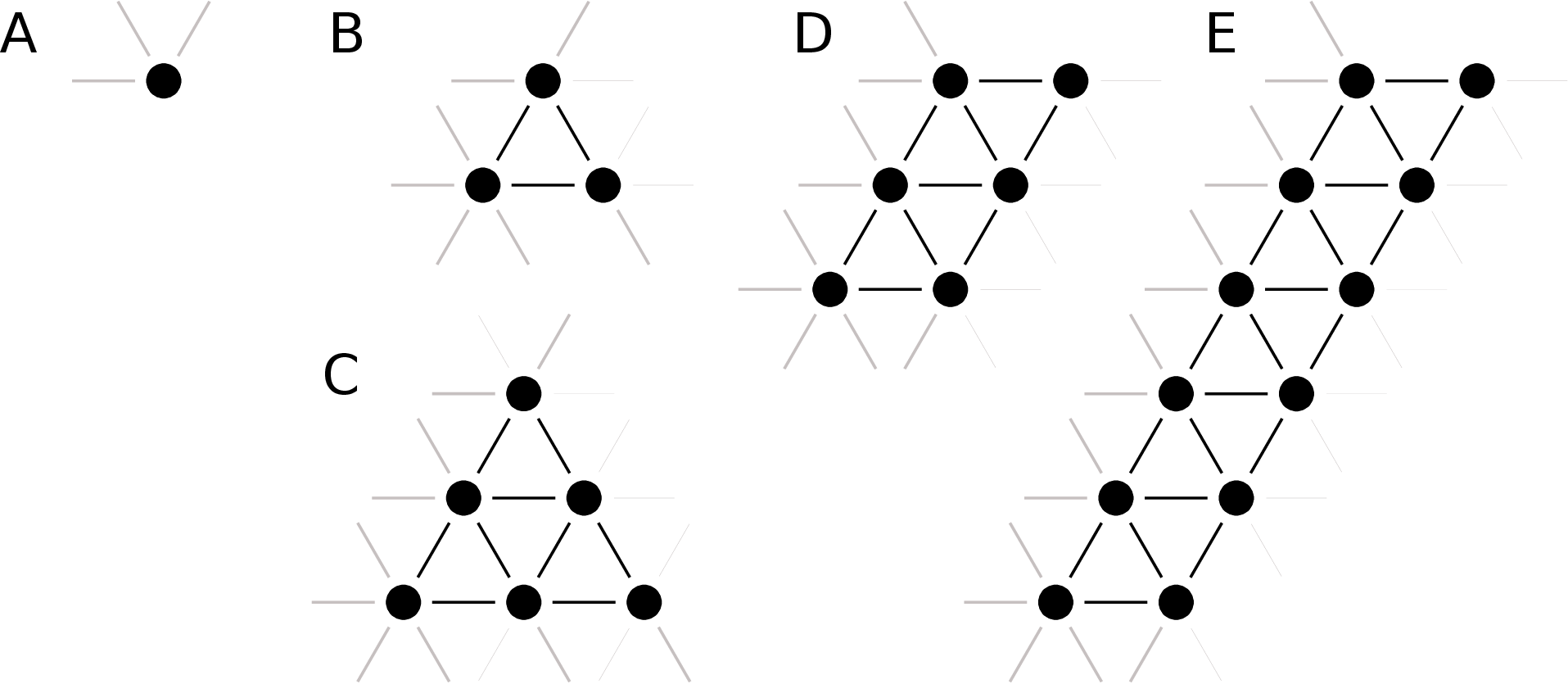}
\caption{Examples of different cluster geometries employed in the simulations: single site MF cluster (A) with $nc=1$ and $N_c=36$ MF Gutzwiller sites, triangular clusters (B with $nc=3$, and C with $nc=6$) with $N_c=2$ and square tessellations (D with $nc=6$ and E with $nc=12$ sites) with $N_c=3$ independent clusters.}
\label{fig:cmf_graphs}
\end{figure}
By systematically scaling the cluster size $nc$, one may obtain quantitative exact results for 2D-systems when comparing to more extensive QMC studies~\cite{Dynamical critical, zhaojize, Singh2018}. Therefore, we consider different cluster sizes and tessellations of the triangular lattice to examine the stability of our findings. In Fig.~\ref{fig:cmf_graphs}, we exemplify different cluster sizes employed in the study.

In practice, we determine the self-consistent solutions $\rho_i^\sigma$ and $\Psi_i$ by iterative calculation of the ground state of the cluster-system until the mean-fields have converged. While for small cluster sizes such as Fig.~\ref{fig:cmf_graphs}(E) the ground-state of the cluster may be obtained by exact-diagonalization techniques, for the larger cluster sizes we employ the DMRG simulation. Following Ref.~\cite{clusterDMRG} in this MF scheme already a low number of matrix-states is sufficient. In the present examples we choose $30$ and $60$ states for $nc=12$ site cluster results shown below.

The solid or density wave orders denoted by  $\Delta\rho^a$, $\Delta\rho^{\sigma}$, and $\Delta \Psi$ are defined by
\be \label{eq:2}
\delta A  =\frac{1}{nc}\sum_{i\in c} |A_i-\bar{A}|,~ \bar{A}=\frac{1}{nc}\sum_{i\in c}\bar A_i.
\ee
We also define the total excitation $\rho=\rho^a+\rho^\sigma$~\cite{SMI}, and  $\Delta \rho = \frac{1}{2}({\Delta\rho^a+\Delta\rho^\sigma})$.

\subsection{MF results for the triangular lattice}

\begin{figure}[b]
\centering
\begin{minipage}[t]{.1\linewidth}\raisebox{6.5\height}{(a)}\end{minipage}%
\begin{minipage}{.38\linewidth}
\includegraphics[width=\linewidth,trim={10cm 2cm 10cm 2cm},clip]{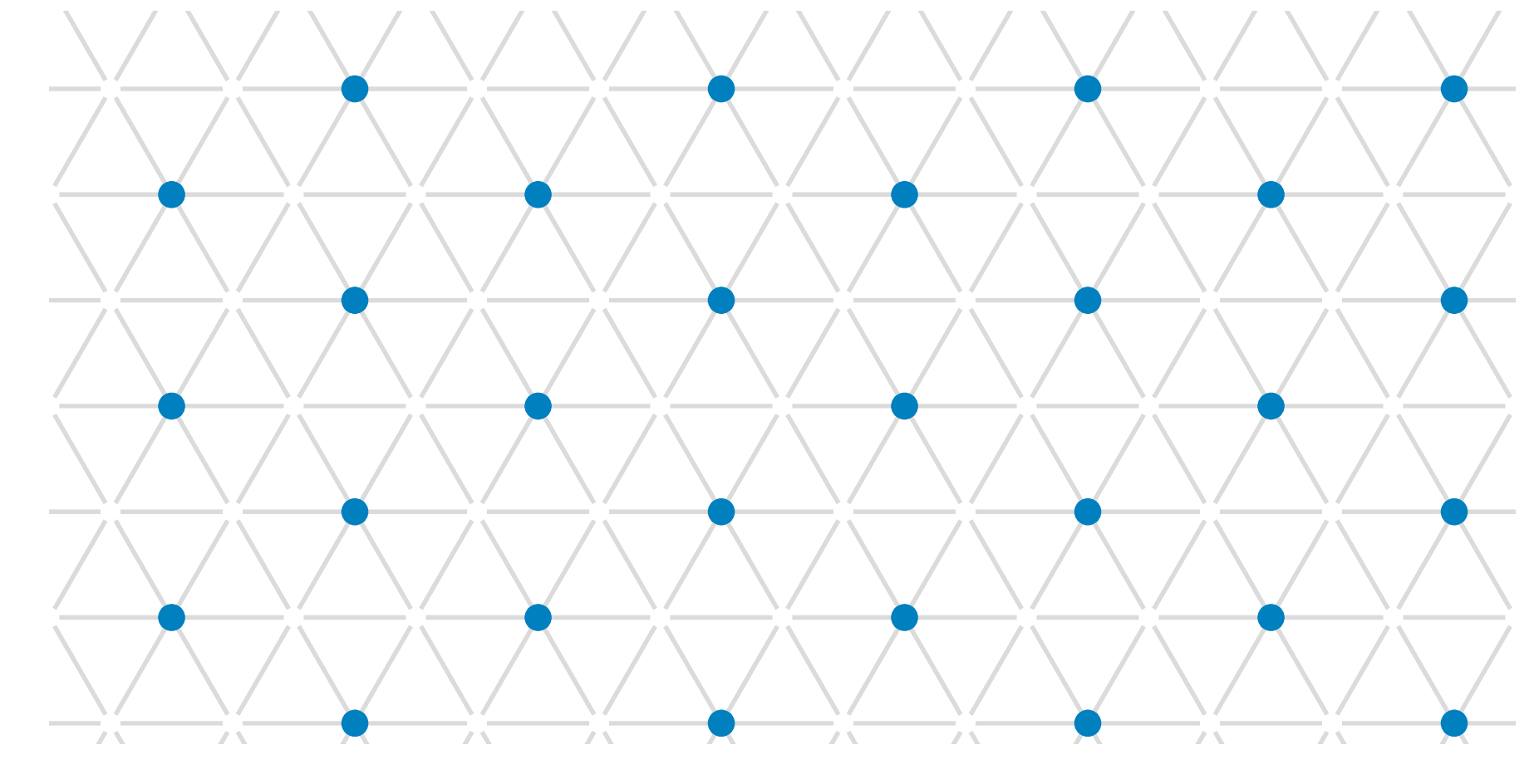}
\end{minipage}
\begin{minipage}[t]{.1\linewidth}\raisebox{6.5\height}{(b)}\end{minipage}%
\begin{minipage}{.38\linewidth}
\includegraphics[width=\linewidth,trim={10cm 2cm 10cm 2cm},clip]{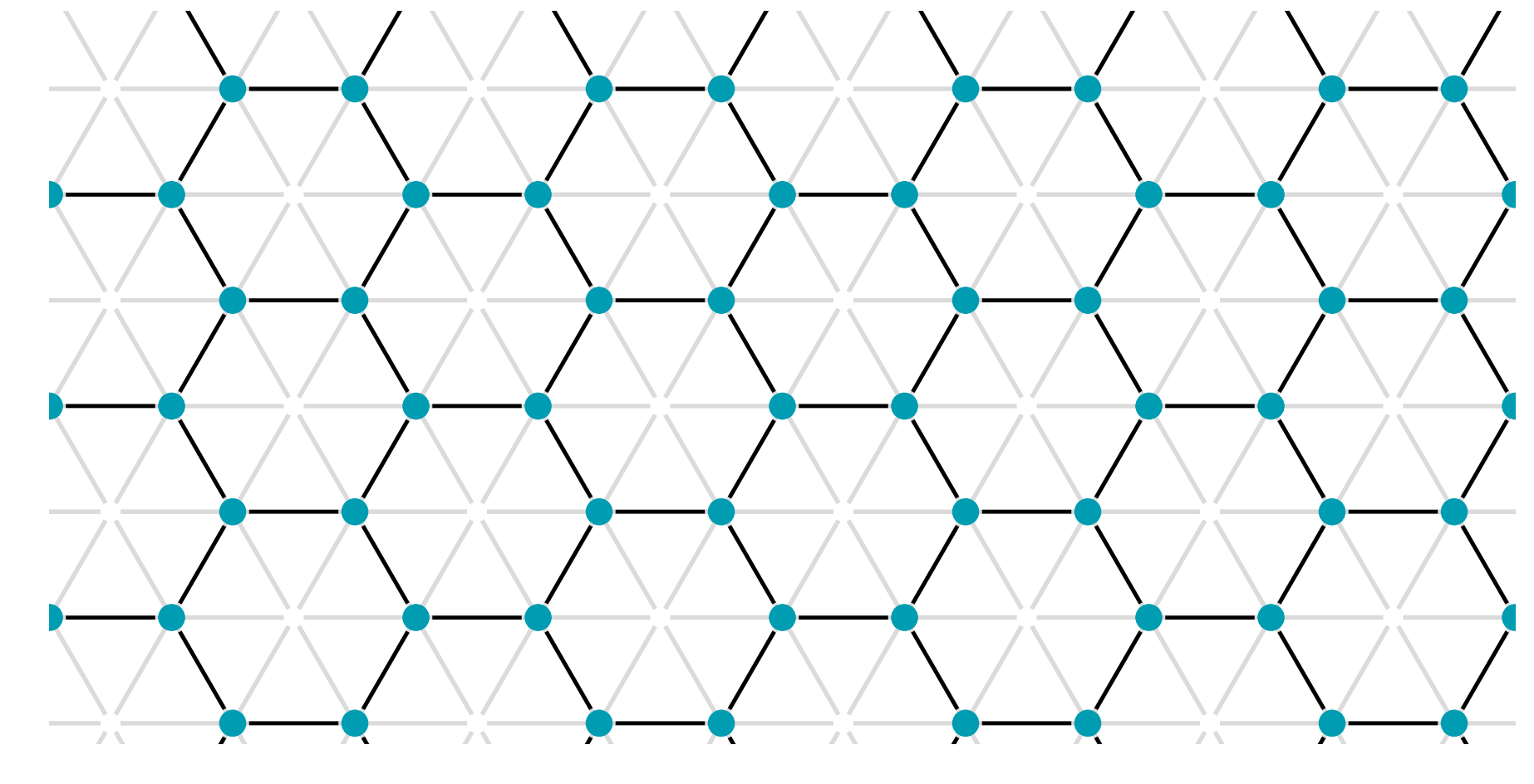}
\end{minipage}
\caption{Sketch of the typical density configurations (blue bullets) for (from 1-site Gutzwiller method) for the (a) DW$_{1/3}$ phase ($t/\beta=0.015$, $\mu/\beta=-0.9$) and (b) the DW$_{2/3}$ phase ($t/\beta=0.015$, $\mu/\beta=-0.7$). The black lines indicate the strength of the nearest-neighbor density correlations $\la n^b_i n^b_j \ra$.}
\label{fig:tri2d_phases_sketch}
\end{figure}
\begin{figure}[t]
\centering
\includegraphics[width=0.48\linewidth]{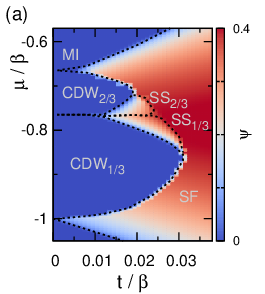}
\includegraphics[width=0.48\linewidth]{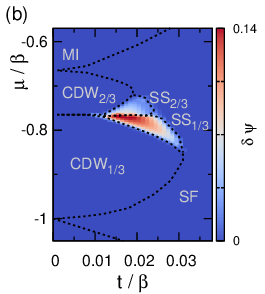}
\caption{Schematic phase diagram of the JCH model as obtained from the Gutzwiller 1-site MF approach ($N_c=36$) in the $\mu-t$ plane for $V/\beta=0.4$. The colors depict the (a) superfluid order parameter $\Psi$ and (b) the supersolid order $\delta \Psi$.}
\label{fig:fig_2dgw_JCH_pd_beta1_V0.4}
\end{figure}

\begin{figure}[b]
\centering
\begin{minipage}[t]{.1\linewidth}\raisebox{6.5\height}{(a)}\end{minipage}%
\begin{minipage}{.38\linewidth}
\includegraphics[width=\linewidth,trim={10cm 2cm 10cm 2cm},clip]{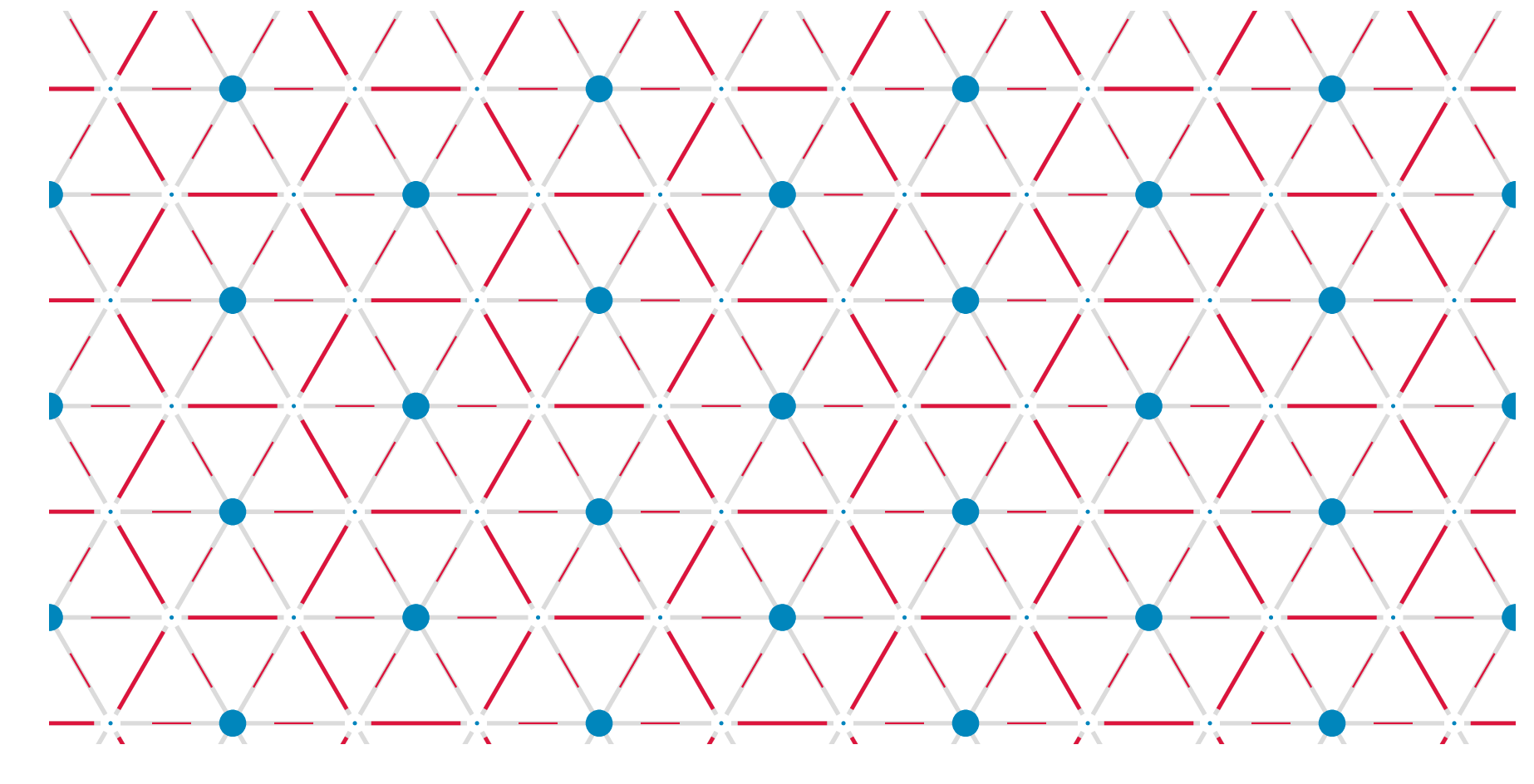}
\end{minipage}
\begin{minipage}[t]{.1\linewidth}\raisebox{6.5\height}{(b)}\end{minipage}%
\begin{minipage}{.38\linewidth}
\includegraphics[width=\linewidth,trim={10cm 2cm 10cm 2cm},clip]{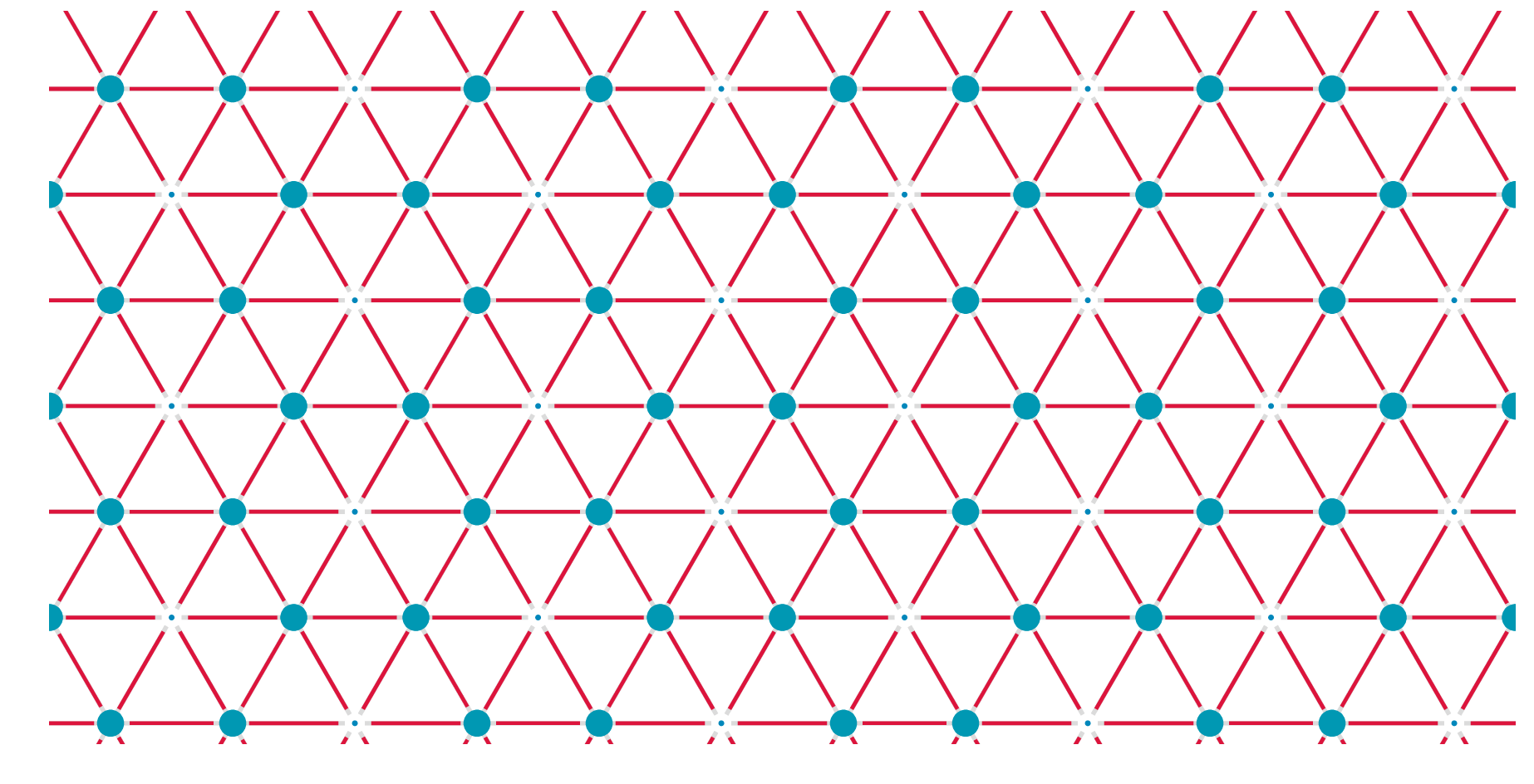}
\end{minipage}
\caption{Sketch of the typical density configurations (blue bullets) for (from 1-site MF method) for the (a) SS$_{1/3}$ phase ($t/\beta=0.022$, $\mu/\beta=-0.79$) and (b) the SS$_{2/3}$ phase ($t/\beta=0.022$, $\mu/\beta=-0.74$). The red lines indicate the strength of the bond kinetic energy $|\la a^\dagger_i a_j \ra|$.}
\label{fig:tri2d_phases_SS_sketch}
\end{figure}

In the limit $t = 0$ and large $\beta\gg V$, where Eq.~\eqref{aBH} is valid, the energy per unit cell is
\begin{equation}
\begin{aligned}
E_{\Delta}=&\sum_{i}[-\beta -\mu] n_i^b +\frac{3 V}{4} (n_A^\sigma n_B^\sigma +n_B^\sigma n_C^\sigma+n_C^\sigma n_A^\sigma),
\end{aligned}
\end{equation}
where $n_A^\sigma, n_B^\sigma$ and $n_C^\sigma$ are the number of excitations on alternating sublattice sites $i=A, B$, and $C$.
In analogy to the zigzag ladder case, by increasing $\mu$, the density $\rho$ will undergo platforms with values of $1/3$, $2/3$ with solid patterns shown in Fig.~\ref{fig:tri2d_phases_SS_sketch}. A solid phase at half filling is not present in the 2D case. In a grand-canonical ensemble, the two solid phases exhibit a direct transition at $t \rightarrow 0$ with all states where $1/3<\rho<2/3$ being macroscopically degenerate in this classical limit. The hardcore BH model on the triangular lattices has been studied in Refs.~\cite{tri2,tri3,pairss} on triangular lattices for a finite hopping $|t|>0$ and the emergence of supersolid phases stabilized by an order-by-disorder mechanism for intermediate fillings $1/3<\rho<2/3$, has been shown.

In Fig.~\ref{fig:tri2d_phases_SS_sketch} we present the results of single site MF simulations for these parameters which show the two solid phases DW$_{1/3}$ and DW$_{2/3}$ for a finite atom-photon coupling $\beta$ in the limit of vanishing and small $t\ll\beta$.
It is important to note, that for $t \rightarrow 0$ as known for the BH limit, the two phases exhibit a first order transition with a macroscopic jump of density between the two platforms.
Furthermore, as shown in Fig.~\ref{fig:tri2d_phases_SS_sketch},  due to the coupling between atoms and photons, the DW$_{1/3}$ and DW$_{2/3}$ phases are not symmetric
with $\mu/\beta=-0.77$, which is an important difference to the
particle hole symmetry of the hardcore bosons on the triangular lattices~\cite{tri2,tri3,pairss}.

Interestingly, as shown in Fig.~\ref{fig:tri2d_phases_SS_sketch}, two SS phases appear between the DW$_{1/3}$ and DW$_{2/3}$ phases with $\delta \Psi \ne 0$. Sketches of the two different SS-patterns are shown in Fig.~\ref{fig:tri2d_phases_SS_sketch}.
Between the two solids, the SS phase can be understood in terms of photon-tunneling breaking the degeneracy between the DW$_{1/3}$ and DW$_{2/3}$ phases. Hence,
the emergence of such a SS phase may be based on an order-by-disorder mechanism as conjectured for BH limit in Refs.~\cite{tri2,tri3,pairss}.

\begin{figure}[tb]
\includegraphics[width=\linewidth]{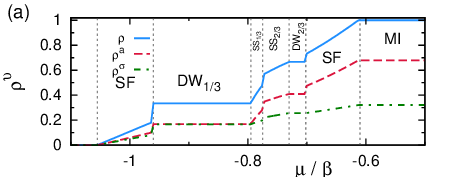}\vskip 0.03cm
\includegraphics[width=\linewidth]{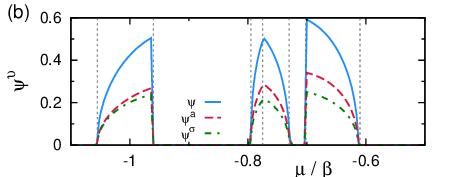}\vskip 0.03cm
\includegraphics[width=\linewidth]{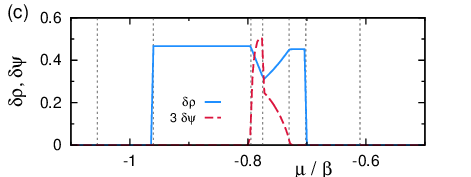}
\caption{CMF method detailed description of the (a) total density $\rho$, and density of photons and excitations $\rho^a$ and $\rho^\sigma$, (b) the SF order $\Psi$, $\Psi^a$ and $\Psi^\sigma$ and the (c) (super)-solid order $\delta \rho$ and $\delta \Psi$ vs $\mu/\beta$ with $V/\beta=0.4$, $t/\beta=0.018$ on the triangular lattices. CMF data for clusters of 3-site triangular shape (B in Fig.~\ref{fig:cmf_graphs}).}
\label{fig:CMFtri3data}
\end{figure}

\subsection{Stability of the mean-field results with the cluster size}

To illustrate the above results with more detail and study the stability of the 1-site MF results, we scan the phase diagram along the lines with different values of $t/\beta$. As an example, we choose $t/\beta=0.018$ in Fig.~\ref{fig:CMFtri3data} obtained now by a 3-site CMF simulation.
Starting at  $\mu/\beta\approx -0.95$ and increasing $\mu/\beta$ to $-0.8$,
the system is in the DW phase with $\rho=1/3$, $\Psi=0$ and $\delta\rho=0.44$.
With a further increase of $\mu/\beta$, the three quantities $\Psi$, $\delta\Psi$ and $\delta\rho$
become nonzero continuously, and the system enters into a SS phase.
Since this SS phase is understood from particle-doping on the DW$_{1/3}$ phase, we therefore denote it as a SS$_{1/3}$ phase.

\begin{figure}[tb]
\includegraphics[width=\linewidth]{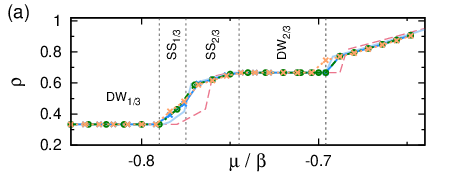}\vskip 0.03cm
\includegraphics[width=\linewidth]{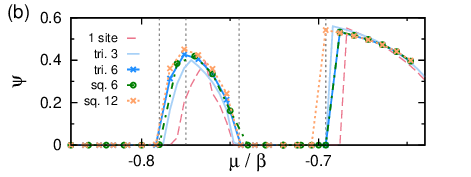}\vskip 0.03cm
\includegraphics[width=\linewidth]{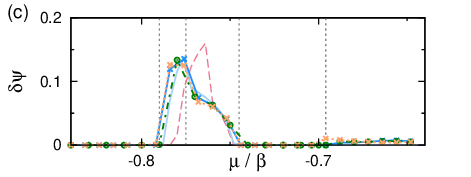}
\caption{Comparison and scaling of the CMF results for various cluster-sizes for (a) the density $\rho$, the (b) SF order $\Psi$ and (c) supersolid order $\delta \Psi$ vs $\mu/\beta$ with $V/\beta=0.4$, $t/\beta=0.015$ on the triangular lattice. The different curves show the 1-site MF simulation (simulation of a $N_c=36$ site-unit-cell), different clusters of triangular shape (B and C in Fig.~\ref{fig:cmf_graphs} with $nc=3$ and $nc=6$ sites, $N_c=2$) and square-parketting (D and E in Fig.~\ref{fig:cmf_graphs} with $nc=6$ and $nc=12$ sites, $N_c=3$).}
\label{fig:compCMF}
\end{figure}

By increasing $\mu/\beta$ to $-0.75$, the quantities $\rho$ and $\Delta\Psi$ jump to  nonzero values and the system enters into another SS phase. This SS phase could be called SS$_{2/3}$, because it is formed by hole-doping on the solid DW$_{2/3}$.
The phase transition of SS$_{1/3}$ to SS$_{2/3}$ is first order. This is consistent with previous work~\cite{Jap,wang2013}, in which the two SS phases of hardcore bosons take place according to first-order transitions. By continued increase of $\mu/\beta$ to $-0.7$, it is obvious that $\Psi$ becomes nonzero,
which means that the first-order DW$_{2/3}$-SF phase transition takes place.

In previous works, the nearest~\cite{8} or next to nearest~\cite{next} repulsive interactions are the necessary conditions for the formation of the supersolid.
It should be noted that in our model, even though there is no interaction between photons, the light supersolid emerges,
because the repulsion of atoms will cause the effect of repulsion between photons due to atom-photon coupling.
This effected atom-photons can be verified by calculation of the expectation values of
$\sigma_i^{\dagger}a_i$ or $\sigma_i a_i^{\dagger}$, which is nonzero even the system is in the DW$_{1/3}$, DW$_{2/3}$, MI($\rho=1$) phases,
where $\Psi=0$ (not shown). The behaviours of $\langle\sigma_i^{\dagger}a_i \rangle \ne0$ and $\langle\sigma_i^{\dagger}\sigma_{i+1}^{\dagger}\rangle \ne0$ mean that a type of fluctuation with transitions  between  atom excitations and photons, remains in the system.

Within the given range of parameters and observables our results remain stable of the MF results with the cluster size. As exemplified in Fig.~\ref{fig:compCMF} already the 1-site MF approach may qualitatively reproduce the main features of the phase diagram, compared to the CMF results. The 3-site and 6-site CMF results already give quantitative good agreement and the 12-site data mainly overlaps with the 6-site results. Interestingly, we observe only a little difference between the different 6-site tessellations of Fig.~\ref{fig:cmf_graphs}~(C) and (D) employed in the CMF simulation. In particular, the size of the SS-phases increases slightly with the cluster size, which indicates that this phase may be stable in the 2D triangular JCH model.
We want to note that our CMF findings should be complemented by a QMC study of the model to analyze the stability and properties of the supersolid phases in more detail. However, this goes beyond the scope of this work and will be discussed elsewhere.

\section{Discussion and conclusions}
\label{sec:con}
Through a systematic study of the extended JCH model on triangular lattices, we find that the light supersolid is stable in coupled cavities in
the thermodynamic limit even when the photon hopping term is considered.
On the triangular lattices,
in the mean-field and DMRG  phase diagrams, the area of the supersolid is relatively wider than that in the one dimensional bipartite lattice, which is helpful
for experimental detection.
For both the hardcore and softcore JCH models, using MF and the DMRG methods, we find the SS phases.

It is worth mentioning that we find a novel pair correlation caused
by the interplay between the atoms in the cavities and the atom-photon interaction.

For the formation of the SS phase of the BH model, the key is the interaction, which is
absent between photons.  However, the solid and SS of photons still can form through the atom-photon coupling, which induces an effective interaction between the photons.

It may be interesting to test the possibility of existence of pair correlation,  beats and supersolid phases in other atom-photon coupled systems in the future.
Our results, obtained by the MF and DMRG methods, will be helpful in the guiding experimentalists in realizing different and novel quantum phase on optical lattices.

\begin{acknowledgments}
We would like to thank T.C. Scott of BlockFint, Bangkok Thailand in helping prepare this document.
S.G. acknowledges support of project no. SA 1031/10-1 of the German Research Foundation (DFG) and the Swiss National Science Foundation under Division II.
WZ is supported by the NSFC under Grant No.11305113.
\end{acknowledgments}



\end{document}